# Better Research Software Tools to Elevate the Rate of Scientific Discovery

## or why we need to invest in research software engineering

Joran Deschamps, Damian Dalle Nogare, Florian Jug

Fondazione Human Technopole, Viale Rita Levi-Montalcini 1, 20157 Milan, Italy

In the past decade, enormous progress has been made in advancing the state-of-the-art in bioimage analysis - a young computational field that works in close collaboration with the life sciences on the quantitative analysis of scientific image data. In many cases, tremendous effort has been spent to package these new advances into usable software tools and, as a result, users can nowadays routinely apply cutting-edge methods to their analysis problems using software tools such as ilastik [1], cellprofiler [2], Fiji/ImageJ2 [3,4] and its many modern plugins that build on the BigDataViewer ecosystem [5], and many others. Such software tools have now become part of a critical infrastructure for science [6].

Unfortunately, overshadowed by the few exceptions that have had long-lasting impact, many other potentially useful tools fail to find their way into the hands of users. While there are many reasons for this, we believe that at least some of the underlying problems, which we discuss in more detail below, can be mitigated. In this opinion piece, we specifically argue that embedding teams of research software engineers (RSEs) within imaging and image analysis core facilities would be a major step towards sustainable bioimage analysis software.

## The maintenance problem and its technical and social causes

Two major determinants for the widespread adoption of a tool are its usability and its maintenance over time [7,8]. The first case is perhaps the more obvious: tools that are difficult to install or use are often neglected by the community of life scientists who could most benefit from them. The second case is perhaps more overlooked, although no less important. Modern bioimage analysis methods are complex, and even the simplest software implementations may depend on dozens of other pieces of scientific software (typically referred to as dependencies). This intricate web of interdependent software packages is, regrettably, extremely brittle, and changes in any one dependency can cause significant changes to be required for existing software tools to continue working properly. Over time, the probability that unmaintained tools are able to still function in up-to-date environments becomes quickly smaller and smaller. Even worse, the time and technical knowledge required to update bioimage analysis software tools is enormous [9], and few people have the know-how to do so.

In a perfect world, all software tools that find active users would be continuously maintained. Unfortunately, there are many reasons why even published software tools with an active user-base are not. Compared to publishing new methods and tools, the maintenance of existing tools typically does not provide much in terms of career advancement for the maintainer. At best, an existing publication might get cited more often, but the advantage of this pales in comparison with a new paper being published. Additionally, the person working on the maintenance of a software tool might not even be an author on the original publication, leaving this important work completely unrewarded by commonly used scientific performance indicators. Finally, maintaining software requires continuity of expertise, which is too easily lost when a main developer switches between labs or moves on to their next career stage. Incentive structures are therefore not aligned with the long-term commitment required to maintain open research software [10], leading to most tools being maintained only for a limited period of time [11].

Another difficulty in maintaining existing software tools is that funding for this kind of work is rare. While there exist both public and private funding mechanisms to support open source projects in general, a recent EU analysis [12] found that the majority of this funding was dedicated to new projects, with comparatively little funding for long-term maintenance of existing ones. There are only a few notable exceptions we know of, where funding bodies have dedicated specific calls to software sustainability (e.g. the Chan-Zuckerberg Initiative [13], the DFG [14] and BMBF (de.NBI), the Virtual Institute of Scientific Software [15], or the Software Sustainability Institute [16]). Still, such funding is currently not available at the level required to build user friendly and better maintained software tools [10].

Currently, not only are we unable to achieve these important goals, we struggle even to keep our best research software engineers employed and offer them a stable career or adequate income. It is therefore not surprising that many of the software engineering activities in academia are carried out by personnel in limited-term positions, such as PhD students, post-docs, or young RSEs on short term staff positions. Since experienced RSEs are highly qualified experts that are also in high demand outside academia, we regularly see talented people leave to work in industry. This makes the previously discussed software maintenance dilemma even more problematic: once such developers run out of time or funding, development is halted, maintenance is abandoned, and the expertise in which the hosting group and institute has invested is lost.

Could the existing academic technology transfer infrastructure be a solution for this problem? Instead of talented RSEs leaving to pursue a career in mainstream technology companies, they could be offered the chance to continue the development and maintenance of research software in the context of a newly incubated startup or spin-off company. There are successful examples, such as Cytomine [17] or KNIME [18], where this has worked, but also many other instances where it has not. The idea of useful software receiving funding by selling licenses is obvious and clean, however it unfortunately often fails due to some combination of the following reasons: *(i)* it is hard to find investors for open software, but closing the code base removes community engagement and slows development and interoperability, *(ii)* the existing codebase is released under a license that makes it hard to commercialize it, *(iii)* many users would rather switch to alternate software tools rather than pay money for licenses. The last point is critical: scientific analysis software is highly specialized and the number of users is, in absolute terms, limited. While some commercial software tools create value by enabling faster, more reproducible, more quantitative, and less biased academic research, the scientists and research groups in the user base may not have the money for software licenses. In addition, analyses performed in commercial software require other research groups to obtain similar licenses to replicate them, creating additional barriers to reproducibility and the dissemination of open science. We therefore believe that research software development should be supported by science funders and institutions.

Despite the undeniable importance of software tools in life science research, being a research software engineer has not yet become a common career path. Individuals that do choose to become RSEs are likely to suffer from job insecurity and will in many cases earn well below equally skilled colleagues working in industry.

But not all hope is lost. There are research groups and scientific institutes that are increasingly investing in their RSEs (a very incomplete set of examples are LOCI at the University of Wisconsin, the EMBL in Heidelberg or Hinxton, the MPI-CBG in Dresden, the Allen Institute in Seattle, or the Human Technopole in Italy).

## Research software engineering in image analysis facilities

We believe that institutes and funding agencies have the power to elevate the usability and sustainability of bioimage analysis software by providing better career opportunities for RSEs. Such change would have a profound impact on the rate of scientific progress even beyond the field of bioimage analysis. More concretely, we postulate that image analysis facilities (or image analysis teams within imaging facilities) are an ideal place for RSEs to conduct their work and foster this potential.

Historically, bioimage analysis has often found fertile ground within imaging facilities, which have naturally been the first to see how important and tightly coupled adequate analysis methods and workflows are to the microscopy work conducted in those facilities. More recently this has led to increased hiring of skilled bioimage analysts in such facilities and even to the creation of dedicated core facilities focused on image analysis (examples can be found, among other places, at Institut Pasteur [19], Institut Curie [20], EMBL Heidelberg [21], Human Technopole [22], or the TU-Dresden [23]).

Key performance indicators (KPI) of image analysis facilities or image analysis teams within imaging facilities are typically metrics like (*i*) the number of interactions with users, (*ii*) the level of satisfaction those users report, (*iii*) the number of completed analysis projects, or (*iv*) the tally of co-authorships or other acknowledgements in resulting publications. The imperative to demonstrate usefulness using such KPIs can, however, leave little space for research software engineering activities. As a consequence, efforts spent on developing new methods and approaches remain circumscribed to the needs of the facility's own users and too rarely impact users more globally. It is up to individual bioimage analysts to devote the extra time required to package an analysis pipeline into a useful tool and release it to the public. There exist examples of methods that were successfully translated into more generally available and popular software tools, such as Trackmate [24,25], MoBIE [26], Labkit [27], CellPose [28], and others, which are broadly used in life-science research. Despite this success, these tools must now be maintained over time, bringing us right back to the above discussion: the difficulty in finding the time, money, and expertise to conduct software maintenance activities for analysis tools that others rely on.

## Giving core facilities a new mission

A tremendous increase in the quality of open research software can be achieved *(i)* by making the development of FAIR (Findability, Accessibility, Interoperability, and Reusability) [29,30] and sustainable software a core mission of image analysis facilities (or analysis teams within imaging facilities) [31], *(ii)* by adding stable RSE positions to teams of bioimage analysts, and *(iii)* by facilitating networking activities between such teams across different facilities. We believe this would lead to more powerful and general-purpose core libraries that would be developed, maintained, and used by a larger group of research software engineers. As a consequence, users would have access to a set of software tools that have a higher degree of stability, interoperability and require less maintenance thanks to shared maintenance efforts, rather than duplicated ones. Additionally, this would mean that more RSEs would be involved in developing such core libraries, which would in turn increase those libraries bus-factor[1] and thereby mitigate many of the software maintenance problems arising from single individuals leaving their jobs. All together, this would result in a more efficient translation from new methods to usable tools that can benefit bench scientists and therefore directly elevate the rate of scientific discovery.

Still, increased hiring of research software engineers alone is not a guarantee for success. It is also imperative to collaborate across team and institute boundaries and to broadly establish modern software development practices. At times, research software is hastily written and “just good enough” to demonstrate a working principle, but not of sufficient quality to easily reuse, share, or build other components stably on top of it. The more general purpose a library or software component is, the more important it is to invest in a good design that enables its reuse. This also means to invest in automated testing and good documentation. Such things take additional time at first, but pay back the investment many times over the lifetime of a software project. As coordinated exchange of best-practices and peer-teaching will become important, we believe that establishing specialized bioimage analysis RSE networks that review and give feedback on each other's activities will be a key component to ensure that global RSE work is as efficient as it can be.

If our argument to include such a network of RSE teams within image analysis facilities is more broadly implemented, we believe that a number of benefits will naturally and directly emerge. In particular, we expect the following benefits to be a direct consequence:

**User-driven requirement assessment:** Bioimage analysts are the interface to bench scientists and microscopists who generate raw image data that needs to be analyzed. As such, they observe every day which analyses are well supported by user-friendly tools, and where gaps exist in the tool landscape. This knowledge, shared with research software engineers and method developers, is key to directing their attention to the most important problems that need addressing.

**Direct functionality and usability feedback throughout the software development life-cycle:** Bioimage analysts are often the first to use new analysis tools and methods and they typically do so in the context of multiple analysis projects, requiring a broader spectrum of features. Last but not least, bioimage analysts not only use software tools but also help others use them on their own. The feedback of bioimage analysts on existing pain points for users can therefore be invaluable for research software engineers who seek to improve their tools. By embedding RSEs within image analysis facilities, a constant feedback stream will help to steer development efforts to where they are most needed and therefore avoid wasting resources.

[1] The bus factor [32] is a commonly used measurement of the risk resulting from information and capabilities not being shared among team members, or such team members not existing in the first place. It is derived from the hypothetical question "How many team members have to get hit by a bus before the project can’t persist"?

**Long-term maintenance of software tools and components:** Close collaboration between analysts and RSEs ensures that software tools and the libraries they depend on stay well maintained and that functionality-impeding bugs will be resolved quickly. Additionally, the distribution of knowledge over multiple people will increase the above mentioned bus-factor, enabling RSE teams to retain expertise even if some members leave.

**Software interoperability and deduplication of effort:** Bioimage analysis and imaging facilities are and need to remain well networked. RSEs within such teams can use these networks to synchronize and deduplicate efforts across teams and institutions. As a consequence, once common libraries are developed and used, software tools will become interoperable, even if not developed collaboratively. This will reduce the cost of research software development, free up much needed RSE time, and most importantly, benefit users who need interoperable software tools for their analyses [7,33].

**Better software engineering and better career paths for RSEs:** Finally, housing RSE teams within imaging or analysis facilities will foster the dissemination of modern software engineering skills [34,35]. Networked RSEs will naturally exchange best-practices. Not only will RSEs themselves benefit from this accumulated knowledge, these skills can also be shared with PhD students and post-docs who develop computational methods and are therefore working at the interface with software tool development. Additionally, more places employing RSEs will automatically create an environment where RSEs can switch between jobs, without having to leave academia, and will create new career opportunities for RSE-inclined students or post-docs [36].

If these statements are so obvious, why are many funders and institutions not investing in such ideas? We believe that multiple factors come together. Software development methods have improved tremendously over the past decades. Dependency management, code versioning tools, test-driven development, continuous integration, platforms such as Github, etc. all contribute to making distributed software development possible. Hence, our ideas would have been much harder to realize in the 1990s than they are now. Additionally, the return on investment is not easily quantifiable, at least not until enough institutions commit to the interconnected research software engineering and maintenance model we propose. Still, we now benefit from the necessary infrastructure required to collect useful KPIs regarding which software and libraries are successful and widely used. For example: reporting the number of new software versions being released each year, their usage through download statistics, the number of users interacting with the developers through public platforms such as image.sc or Github, the number of forks or stars project repositories receive, or more classical indicators such as citations/mentions in the literature. Additional KPIs can be collected based on teaching and consulting events and user feedback, or by estimating the benefit of using certain software tools in projects conducted within a facility (e.g. in terms of enabled projects or estimated overhead time if the analysis was conducted without those tools).

While this strategy should hopefully find application in many places, we are currently implementing such a structure at Human Technopole within NoBIAS, the National Bioimage Analysis Service, which is part of the new National Facility for Data Handling and Analysis. We firmly believe that a strong RSE team is a key ingredient for a truly successful image analysis facility that aims at serving many users in the context of heterogeneous life science projects. As part of this commitment, we have built our facility around a model which integrates an RSE team as a core part of the facility. This team has a broad mandate to support the activities of the facility, including code review and optimization for our analysis pipelines, casting new methods into user-friendly tools that can then be used first by the analysis team and then later also by the facility users and the scientific community. Another key part of the NoBIAS RSEs mandate is the maintenance of previously developed tools and contribution to open source projects created by others. We are aiming at accelerating the rate of scientific discovery not only for our users on campus and within our national user base, but also for life science research more globally. In addition to enhancing the ability of our facility to serve our users, we hope that such a model will function as a proof-of-concept for the ideas outlined in this opinion piece, and as an example to other facilities of the advantages of integrating RSEs into an image analysis facility.

## Acknowledgements

We would like to thank Talley Lambert (Harvard Medical) and Eugenia Cammarota (Human Technopole) for valuable feedback on the manuscript.